\documentclass[aps,pra,groupedaddress,onecolumn,nopacs,11pt,tightenlines,notitlepage,graphics]{revtex4-1} 

\usepackage[english]{babel}
\usepackage[dvips]{graphicx}
\usepackage{indentfirst}
\usepackage{tabularx}
\usepackage[normalem]{ulem}

\begin{document}

\title{Temperature Dependence of the Vacancy Formation Energy in Solid 
	$^4$He}

\newcommand{\orcidauthorA}{0000-0000-000-000X} 

\author{Riccardo Rota}
\affiliation{Institute of Physics, \'Ecole Polytechnique F\'ed\'erale de Lausanne (EPFL), CH-1015 Lausanne, Switzerland}
\email{riccardo.rota@epfl.ch}

\author{Jordi Boronat}
\affiliation{Departament de F\'\i sica, Campus Nord B4-B5, Universitat Polit\`ecnica de Catalunya, 08034 Barcelona, Spain}

\date{\today}




\begin{abstract}
	
We  studied the thermal effects on the behavior of incommensurate solid 
$^4$He at low temperatures using the path integral Monte Carlo method. Below a 
certain temperature, depending on the density and  the structure of the 
crystal, the vacancies delocalize and a finite condensate fraction appears. We 
calculated the vacancy formation energy as a function of the temperature 
and  observed a behavior compatible with a two-step structure, with a gap of 
few K appearing at the onset temperature of off-diagonal long-range order. 
Estimation of the energy cost of creating two vacancies seems to indicate an 
effective attractive interaction among the vacancies but 
the large error inherent to its numerical estimation precludes a definitive 
statement.	

\end{abstract}



\maketitle







\section{Introduction}

Solid $^4$He at very low temperature has been the candidate for achieving the 
supersolid state of matter for long time~\cite{balibar_rev,claudi_rmp}. Its extreme quantum nature and its 
stability at relatively low pressures could fit into the desired target. 
However, the search for a clear signature of it has been extremely elusive despite 
 continued effort. Some years ago there was a big excitement after the 
announcement of a finite superfluid fraction in solid $^4$He observed as a 
non-conventional moment of inertia of a sample of hcp $^4$He grown inside a 
torsional oscillator~\cite{Chan1,Chan2}. Soon after this achievement, it was observed that the 
elastic constants of the crystal change also at similar temperatures~\cite{beamish}. Initially, the elastic anomaly was assumed to be independent of the mass 
decoupling in the torsional oscillator~\cite{west}. However, improvements in the 
experimental setup confirmed finally that the change in the response under 
rotation can be understood in terms of only the elastic anomaly and that the 
superfluid signal was not present \cite{chan_no}.

The interest on the possible supersolid phase of $^4$He has motivated 
intense theoretical studies to better understand the role of defects, as 
point 
vacancies~\cite{Pederiva97,Chaudhuri99,Galli01,Galli03,Galli06,Boninsegni06,Clark08,Cazorla09,Yaros10,Rota_prl} and dislocations~\cite{calvo,koning}, when the limit of zero temperature is reached.
Andreev and Lifshitz~\cite{andreev} were the first to postulate that vacancies 
in solid $^4$He should behave as a Bose gas and thus their presence can induce a finite condensate fraction below some critical temperature. More recently, this scenario has been observed in the microscopic simulation of hcp $^4$He crystals with the path integral Monte Carlo (PIMC) method~\cite{Rota_prl}. This numerical study has shown that, in the regimes of low temperatures where the off-diagonal long-range order (ODLRO) appears, the zero-point motion of $^4$He atoms is particularly large and makes the vacancies delocalized over several lattice sites, giving rise to a supersolid behavior.

However, the possibility of observing vacancy-induced Bose--Einstein condensation 
in solid $^4$He at thermal equilibrium seems very difficult, because these 
defects have a large activation energy, of the order of ten Kelvin, even in the 
limit of zero temperature. Therefore, the fraction of vacancies present in the 
crystal is expected to be exponentially suppressed at the low temperatures 
needed to reach the supersolid regime. However, this evidence does not exclude the 
possibility of vacancy-induced supersolidity in metastable conditions out of 
equilibrium, as recently observed in crystalline samples of $^4$He during pulsed 
vacuum expansion ~\cite{toennies}, or in quantum crystals other than solid 
helium \cite{cinti14}.

Another relevant question about the behavior of vacancies in solid $^4$He 
concerns the effective interaction arising among these defects. Although several 
works point out that vacancies tend to attract each other, it is still unclear 
if this interaction can lead to the formation of bound states between them. Some 
studies suggest that vacancies should clusterize and evaporate out of the crystal 
during the annealing process~\cite{Boninsegni06,Rossi08,Pessoa09}, but other 
works do not see any evidence of the clusterization of vacancies inside bulk 
$^4$He ~\cite{Lutsyshyn10}. This discrepancy can be due to the finite-size 
effects which unavoidably affect microscopic simulations and therefore do not 
allow to provide a clear answer about the behavior of the vacancies in a real 
macroscopic system.

The aim of the present work was to study the vacancy formation energy $E_v$ in 
quantum solids as a function of its temperature. One should expect, indeed, that 
the appearance of off-diagonal long range order in incommensurate solid $^4$He 
should present some characteristic signal in the energetic properties of the 
system, similar to what happens in liquid $^4$He across the transition from 
the normal to the superfluid phase, where the derivative of the energy with 
respect to the temperature (specific~heat) presents a singularity at the 
critical temperature. Our results show the emergence of two distinct regimes of 
temperature where $E_v$ assumes different values, in crystals with different 
lattice structures and different densities. Remarkably, the temperature at which 
the behavior of $E_v$ changes corresponds to that at which a finite condensate 
fraction appears in the crystal.

The rest of the paper is organized as follows. In Section~\ref{sec:methods}, we briefly discuss the PIMC method used in the simulations of solid $^4$He with (incommensurate solid) and without (commensurate solid) vacancies. In Section~\ref{sec:results}, we report the results for the vacancy formation energies as a function of the temperature. Finally, in Section~\ref{sec:conclusions}, we summarize the main conclusions of the work.

\section{Method}
\label{sec:methods}

The PIMC method is nowadays a standard tool to study quantum fluids and solids at finite temperature. It estimates, in a stochastic form, the thermal density matrix of an $N$ particle system~\cite{CeperleyRev}. As is well known, the partition function 
\begin{equation}
\label{PartitionFunction} 
Z = \textrm{Tr} \left( e^{-\beta\hat{H}} \right) = \int dR \langle R \vert e^{-\beta\hat{H}} \vert R \rangle 
\end{equation}
allows for a full microscopic description of the properties of a given
system with Hamiltonian \linebreak $\hat{H} = \hat{K} + \hat{V}$ at a temperature $T =
(k_B\beta)^{-1}$ (we use the position basis $\vert R
\rangle = \vert {\bf r}_1, \ldots {\bf r}_N \rangle$, with $N$
the number of particles). In quantum systems, the noncommutativity of the
kinetic  and
potential energy operators (respectively, $\hat{K}$ and $\hat{V}$) make
impractical a direct calculation of $Z$ using its definition~in Equation (\ref{PartitionFunction}).

PIMC  uses the convolution property of the thermal
density matrix $\rho(R,R';\beta)=\langle R \vert e^{-\beta\hat{H}} \vert R'
\rangle$ to rewrite the partition function as
\begin{equation}
\label{Convolution} 
Z = \int \prod_{i=0}^{M-1}dR_i \,
\rho(R_i,R_{i+1};\varepsilon) \ , 
\end{equation}
with $\varepsilon=\beta/M$
and the boundary condition $R_M = R_0$. For sufficiently large $M$, we
recover the high-temperature limit of the thermal density matrix where 
the kinetic  and potential parts factorize (Primitive Approximation). 
Ignoring the quantum statistics of
particles, the distribution law appearing in Equation~(\ref{Convolution}) is
positive definite and can be interpreted as a probability distribution
function which can be sampled by standard metropolis Monte Carlo methods.

The number $M$ of convolution
terms (beads) necessary to reach the convergence of Equation~(\ref{Convolution})
to the exact value of $Z$ is inversely proportional to the temperature of
the system. This means that, when approaching the interesting quantum
regime at very low temperature, $M$ increases quickly, making simulations hard,
if not impossible, due to the very low efficiency in the sampling of the
long chains involved.
It is therefore important to work out high-order
approximation schemes for the density matrix, able to work with larger
values of $\varepsilon$. The approximation we use in this work is called
Chin Approximation (CA)~\cite{Sakkos09}. CA is based on a fourth-order
expansion of the operator $e^{-\beta\hat{H}}$ which makes use of the double
commutator $[[\hat{V},\hat{K}],\hat{V}]$, this term being related to the
gradient of the interatomic potential.

In the case of bosons such as $^4$He, the indistinguishability of
particles does not change the positivity of the probability distribution in
Equation~(\ref{Convolution}) and the symmetry of $\rho(R,R';\beta)$ can be
recovered via the direct sampling of permutations. We use the Worm Algorithm
(WA)~\cite{BoninsegniWorm} which samples very efficiently the permutation
space. This technique  works in an extended configuration space,
given by the union of the ensemble $W$, formed by the usual closed-ring
configurations, and the ensemble $G$, which is made up of configurations
where all the polymers but one are closed. 
The $G$-configurations can be used to compute
off-diagonal observables, such as the one-body density matrix $\rho_1({\bf
	r}_1,{\bf r}_1')$, which~is one of the main goals of the present work. 

The PIMC approach can be easily extended to simulate Bose systems at strictly 
zero temperature, in the so-called Path Integral Ground State (PIGS) method 
\cite{sarsa00}. Indeed, the thermal density matrix $\rho(R,R';\varepsilon)$ can 
be seen as an imaginary-time  propagator and can be used to 
project a trial wave function $\Psi_T(R)$ onto the real ground state $\Psi_0(R)$ 
of the many-body system, according to the formula
\begin{equation}
\Psi_0(R_M) = \int \prod_{i=0}^{M-1}dR_i \,
\rho(R_i,R_{i+1};\varepsilon) \Psi_T(R_0) \ .
\label{eq:PIGS}
\end{equation}

With this assumption, the ground state average of the physical observables can 
be written in terms of a multidimensional integral, which presents a term 
equivalent to the probability distribution appearing in Equation~(\ref{Convolution}). 
The only requirement for the trial wave function $\Psi_T$ is to satisfy the 
symmetry condition imposed by the Bose statistics of the quantum many-body 
system \cite{rossi09,rota_pre}. In this work, we~consider an uncorrelated trial 
wave function $\Psi_T (R) = 1$. Moreover, we make use of the WA also in the 
simulation of solid $^4$He at zero temperature even if it is not strictly 
necessary, as it allows for a better sampling of the probability distribution 
and it provides a simple strategy to guarantee the right normalization of the 
one-body density matrix $\rho_1({\bf r}_1,{\bf r}_1')$ \cite{rota_pre}.


\section{Results}
\label{sec:results}

We have carried out a series of PIMC simulations of solid $^4$He at densities close to the melting point and for temperatures $T\leq 2$ K, with the main goal of determining the thermal evolution of a quantum crystal with point defects. In particular, we have simulated systems of $N$ and $N-1$ particles interacting with the Aziz potential \cite{aziz}, inside a cubic simulation box with periodic boundary conditions. The number of atoms $N$ is chosen to make the crystal geometry commensurate with the simulation box: we have chosen $N = 108$ for the fcc lattice and $N = 128$ for the bcc lattice. In these simulations, we focus our attention in the calculation of two observables: the one-body density matrix $\rho_1$ and the activation energy of a vacancy $E_v$.

The one-body density matrix (OBDM) is defined as 
\begin{equation}
\rho_1({\bf r}_1,{\bf r}'_1) = \frac{1}{\rho Z} \int d{\bf r}_2 \ldots  d{\bf 
	r}_N \rho(R,R';\beta) \ ,
\end{equation}
where the two configurations $\vert R \rangle = \vert {\bf r}_1, {\bf r}_2 
\ldots {\bf r}_N \rangle$ and $\vert R' \rangle = \vert {\bf r}'_1,  {\bf r}_2, 
\ldots {\bf r}_N \rangle$ differ only for the coordinate of one particle and 
$\rho$ is the density of the system. In PIMC simulations, $\rho_1({\bf r}_1,{\bf r}'_1)$ can be computed by mapping the system of $N$ quantum particles on a classical system made of $N-1$ ring closed polymers and one linear open polymer and by building the histogram of the distances between the extremities of the open chain occurred during the Monte Carlo sampling \cite{CeperleyRev}. Moreover, from the long-range behavior of the OBDM, it is possible to infer the condensate fraction $n_0$ of the system, according to the~formula
\begin{equation}
n_0 = \lim_{|{\bf r}_1-{\bf r}'_1| \to \infty} \rho_1({\bf r}_1,{\bf r}'_1) \ .
\end{equation}

{It is important to notice that, despite the finite-size effects which unavoidably affect PIMC simulations, this approach allows for very accurate estimations of $n_0$ in liquid $^4$He, which are in good agreement with experimental measurements over a wide range of pressures and temperatures~\cite{CeperleyRev,rota_jltp,Diallo12}.

The activation energy 
of a vacancy \cite{gillan}
\begin{equation}
E_v = E  ( N-1 , (N-1/N)  V ) - \frac{N-1}{N} \, E(N,V) \ ,
\label{enervacan}
\end{equation}
where $E(N,V)$ is the total energy of the system made up of $N$ particles inside 
a box of volume $V$, is~obtained as the difference between the total energy of 
the crystal presenting a vacancy and that of the commensurate crystal. Notice 
that, in the calculation of the incommensurate system, we rescale the dimensions 
of the simulation box to have the same density as in the commensurate 
solid: thus, our estimation for $E_v$ has to be intended as the vacancy 
activation energy at constant density. Moreover, as $E_v$ is obtained as a 
difference between two energies which scale with $N$, we have restricted our 
simulations to not very large systems to keep the statistical errors 
under control.

\begin{figure}
	\begin{center}
		\includegraphics[width=0.45\linewidth,angle=0]{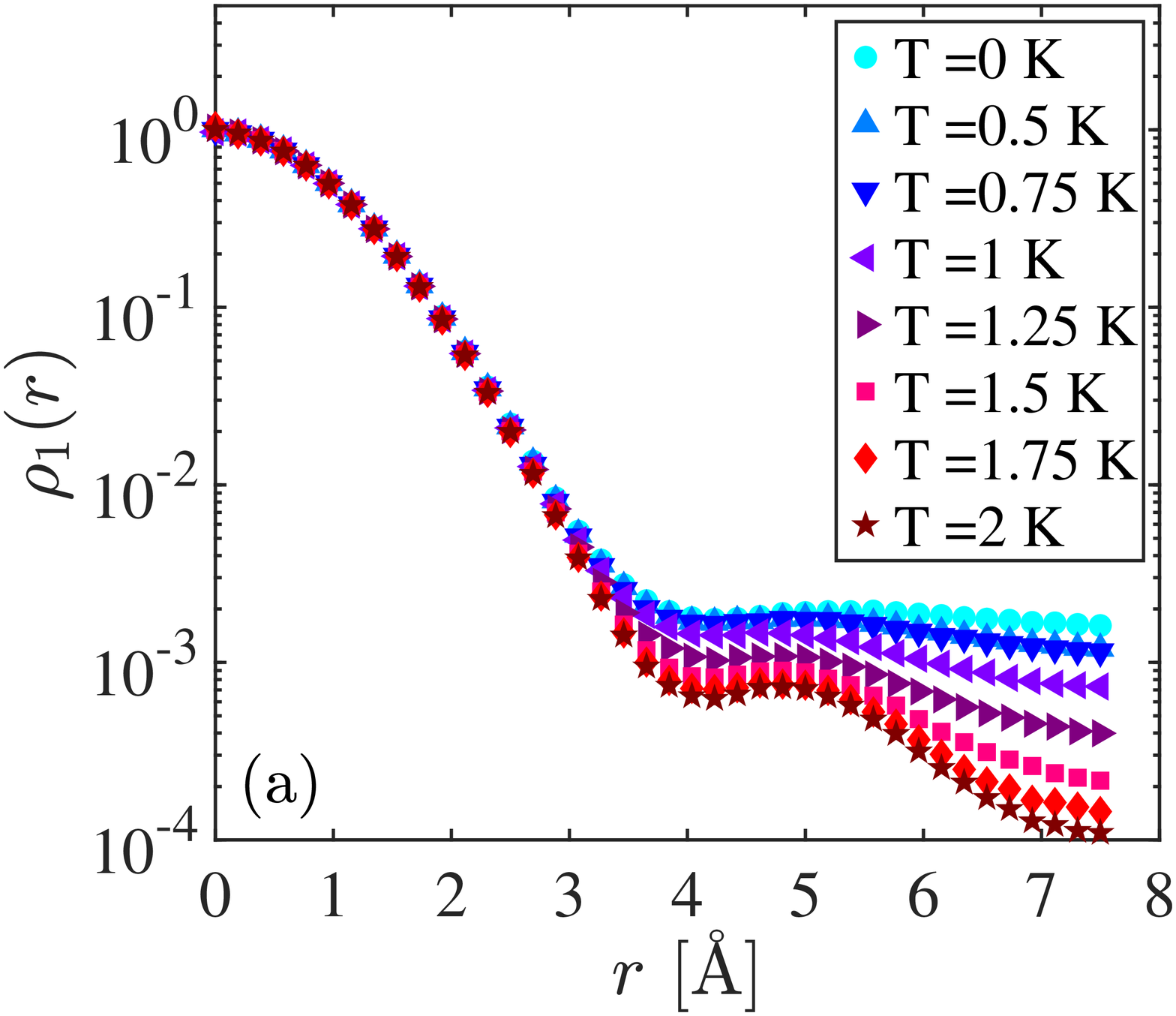}
		\includegraphics[width=0.45\linewidth,angle=0]{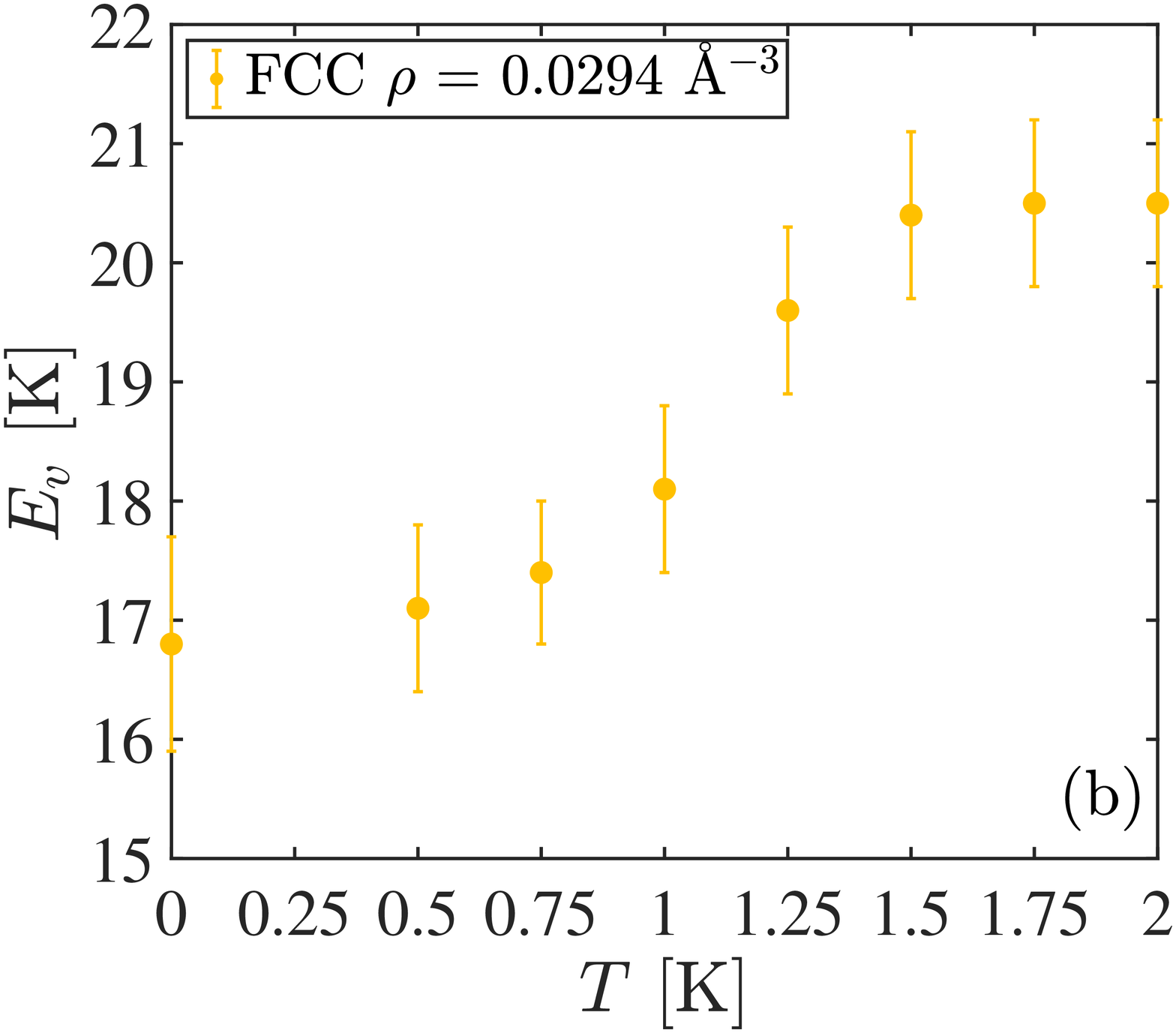}
	\end{center}
	\caption{Results for a fcc crystal at density $\rho=0.0294$ \AA$^{-3}$: (a) one-body density matrix $\rho_1(r)$ as a function of the interatomic distance $r$, for different values of the temperature $T$ (error bars are below symbol size); and (b) vacancy formation energy $E_v$ as a function of the temperature.}
	\label{fcc1}
\end{figure}

In Figure~\ref{fcc1}, we show results corresponding to the fcc crystal at the 
density $\rho=0.0294$ \AA$^{-3}$. Looking at the one-body density matrix 
$\rho_1(r)$ results in Figure \ref{fcc1}-(a), we can see the evolution with temperature. When $T$ is 
large ($T \ge 1.5$ K), we observe an exponential decay which coincides with the 
behavior of $\rho_1(r)$ of the commensurate crystal at any temperature and 
corresponds to absence of Bose--Einstein condensation. 
At these temperatures, the vacancy is localized and the permutations between 
particles are very rare events.
Consequently, the vacancy in this regime behaves in a 
quasi-classical way. When the temperature decreases, we notice some deviations 
in the behavior of $\rho_1(r)$ at large distances until we reach a regime of 
small temperature ($T \le 0.75$ K) where the one-body density matrix shows 
clearly the emergence of ODLRO, pointing to a supersolid behavior where the 
vacancy is delocalized and induces a finite superfluid fraction in the crystal. 
For this case, the condensate fraction at zero temperature is $n_0 = (1.5 \pm 
0.2) \times 10^{-3}$, which is consistent with the variational calculations presented in Ref. \cite{Galli01}. In Figure~\ref{fcc1}-(b), we report results for the vacancy formation energy as a function of the temperature. 
For $T \le 0.75$ K, i.e., where the system presents ODLRO. we find a value $E_v \simeq 17$ K which is almost independent of $T$. Our numerical results are in agreement with the previous PIGS calculation of Ref. \cite{Galli03}. At higher temperature, we notice at first a steep increase of $E_v$ with $T$ and then a regime where again $E_v$ depends very weakly on $T$: for $T \ge 1.5$ K, i.e., where $\rho_1(r)$ decays exponentially, we estimate $E_v \simeq 20$ K. This behavior shows the emergence of two different phases, according to the temperature of the crystal, and indicates that the delocalization of the vacancies implies a reduction in their activation energy of about 3 K.

It is interesting to notice that Clark and Ceperley  \cite{Clark08} claimed that turning on the Bose statistics on a PIMC simulation of solid $^4$He does not have any significant effect in commensurate crystals but leads to reduction of the total energy of the system in crystals with a vacancy. This result is plausible with our hypothesis that the delocalization of the vacancies at low temperature, which~is allowed by the frequent permutations among $^4$He atoms, is at the origin of the behavior of $E_v(T)$ shown in Figure~\ref{fcc1}.

\begin{figure}
	\begin{center}
		\includegraphics[width=0.45\linewidth,angle=0]{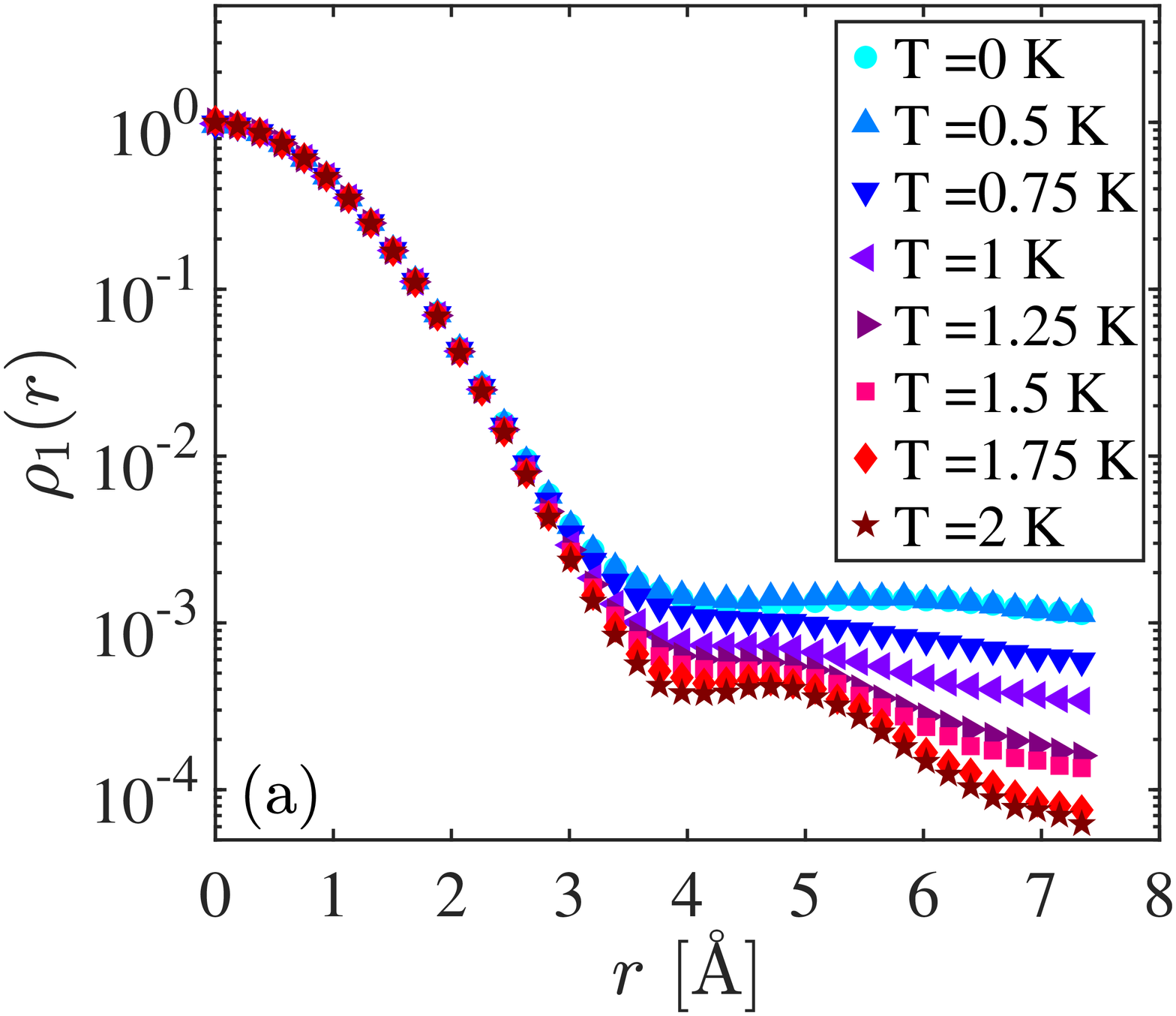}
		\includegraphics[width=0.45\linewidth,angle=0]{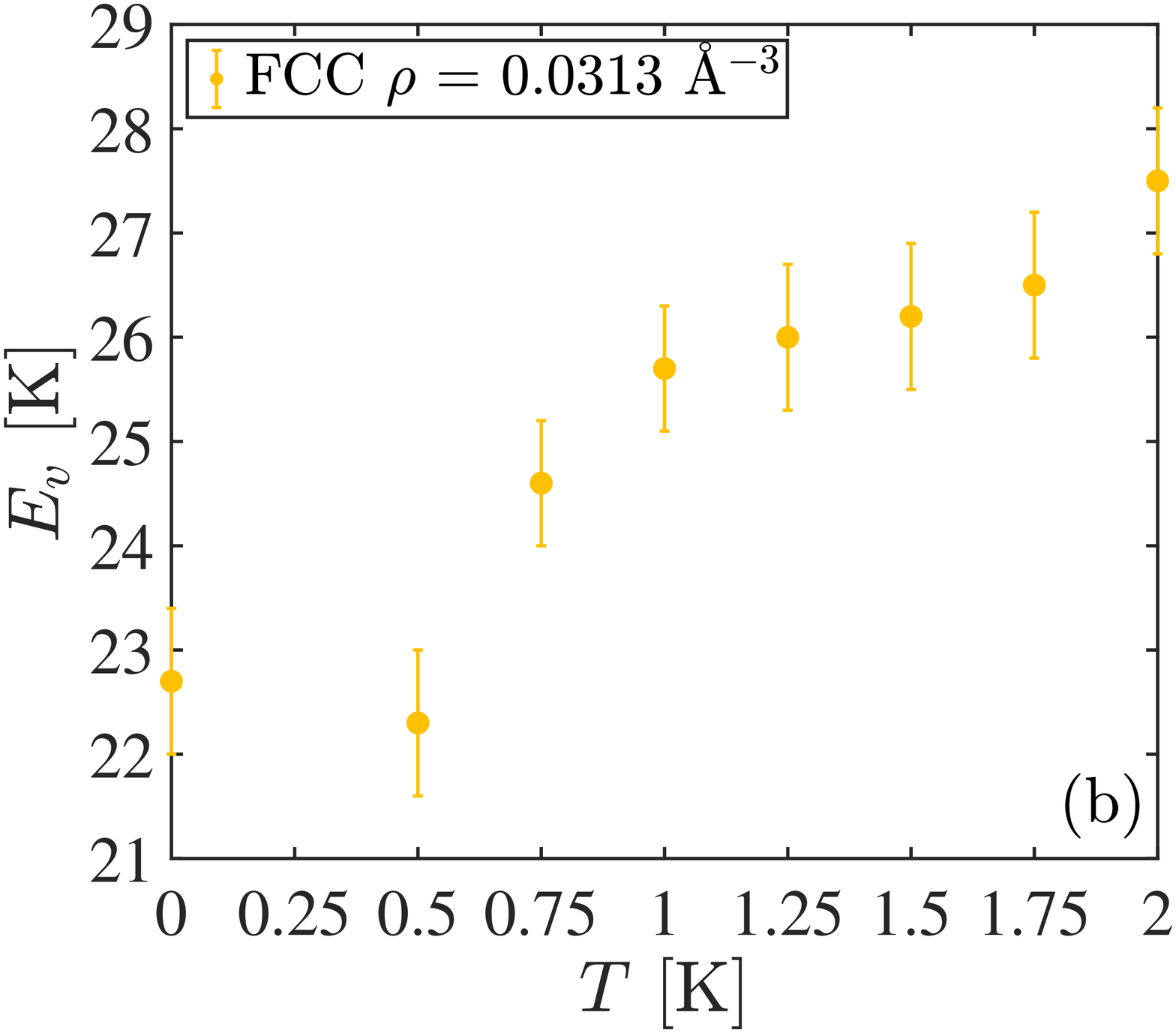}
	\end{center}
	\caption{Results for a fcc crystal at density $\rho=0.0313$ \AA$^{-3}$: 
		(a) one-body density matrix $\rho_1(r)$ as a function of the interatomic distance $r$, for different values of the temperature $T$ (error bars are below symbol size); and (b) vacancy formation energy $E_v$ as a function of the temperature.}
	\label{fcc2}
\end{figure}

To investigate the effect of the density of the system and the 
lattice geometry, we  performed the same calculation in different 
configurations of the crystal. The results obtained for a fcc lattice at a 
higher density, $\rho=0.0313$ \AA$^{-3}$, are shown in Figure~\ref{fcc2}. In this 
case,  clear evidence of a finite condensate fraction is found only for $T \le 
0.5$ K and the activation energy of the vacancy is larger than in the previous 
case. The increase of $E_v$ with the density agrees with 
previous results at zero temperature \cite{Pederiva97}. However, despite 
these small quantitative differences, its qualitative behavior is 
similar to what is shown in Figure~\ref{fcc1} and confirms the evidence that the 
activation energy of a delocalized vacancy is smaller than that of a localized 
one. At this density, the \textit{gap} between the energies in the two 
phases is about 5 K.

\begin{figure}
	\begin{center}
		\includegraphics[width=0.45\linewidth,angle=0]{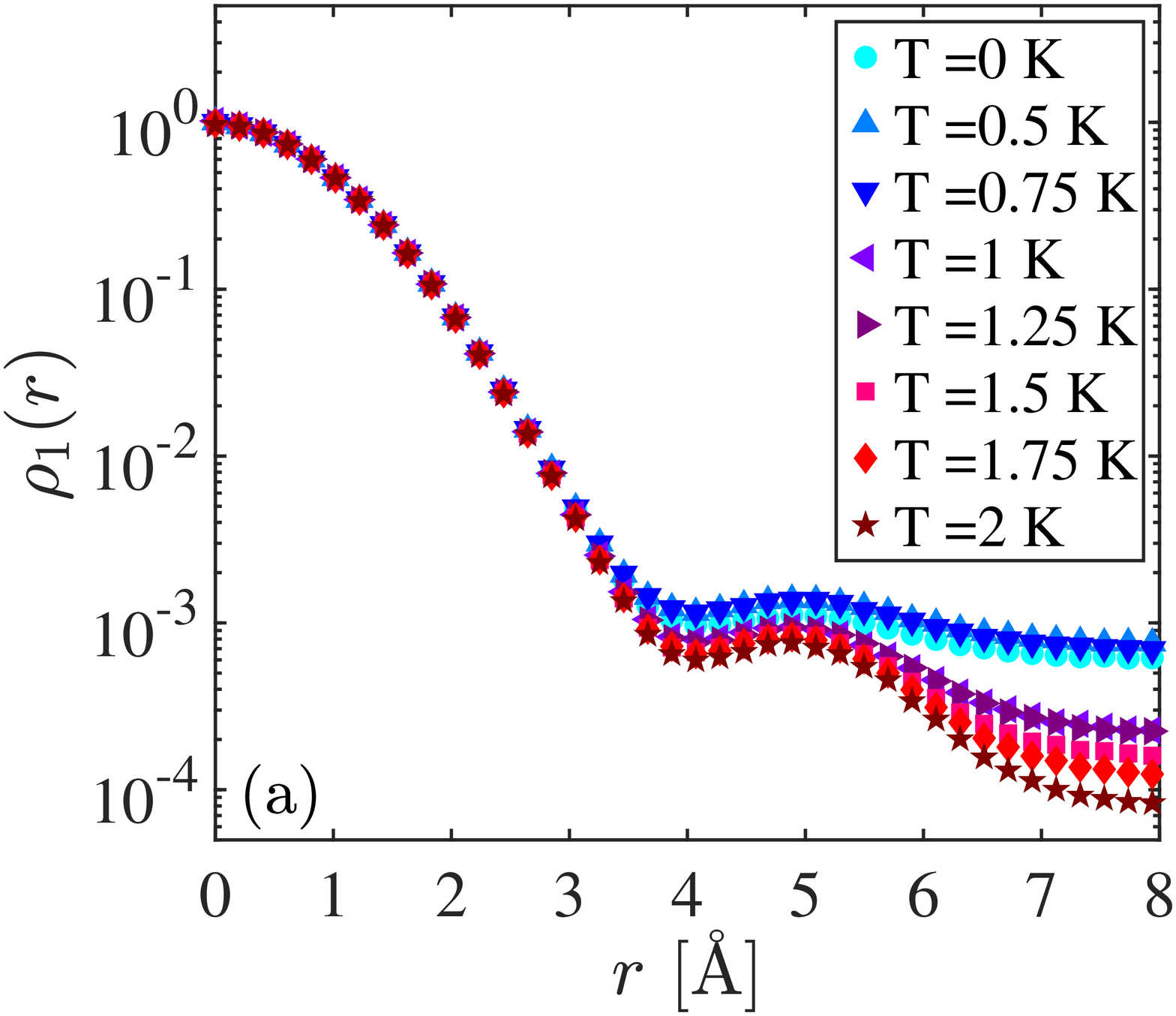}
		\includegraphics[width=0.45\linewidth,angle=0]{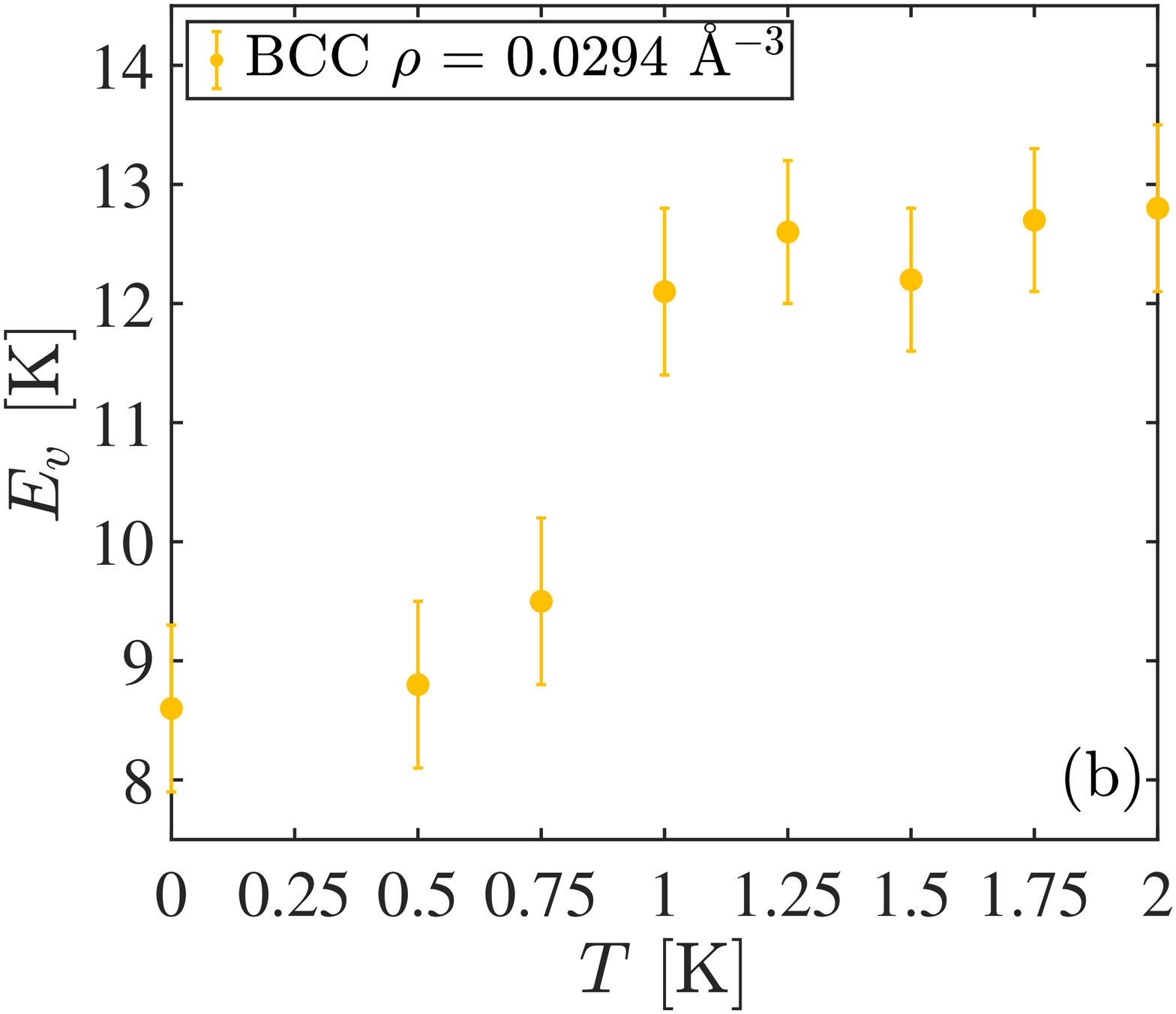}
	\end{center}
	\caption{Results for a bcc crystal at density $\rho=0.0294$ \AA$^{-3}$: 
		(a) one-body density matrix $\rho_1(r)$ as a function of the interatomic distance $r$, for different values of the temperature $T$ (error bars are below symbol size); and (b) vacancy formation energy $E_v$ as a function of the temperature.}
	\label{bcc1}
\end{figure}

In Figure~\ref{bcc1}, we show the results obtained for a bcc crystal with $N=128$ 
sites at a density $\rho=0.0294$ \AA$^{-3}$. The quantitative differences between the activation energy of a vacancy in bcc and fcc crystals have  already been observed in Ref. \cite{Chaudhuri99} and can be ascribed to the less packed structure of the bcc lattice. Nevertheless, we notice once again a clear two-step behavior in the curve 
for $E_v(T)$: for~$T \le 0.75$ K, where $\rho_1(r)$ shows clearly the presence 
of a condensate fraction $n_0 = (7.0 \pm 0.8) \times 10^{-4}$, the activation 
energy of the vacancy is $E_v \simeq 8.5$ K; instead, in the regimes of large 
temperatures, a~larger value $E_v \simeq 12.5$ K is found.

\begin{table}
	\centering
	\begin{tabular}{ccccc}
		\hline
		\hline
		\boldmath{$T$}  & &  \boldmath{$E_v$} &  & \boldmath{$E_{2v}$}  \\
		\hline
		0.5   & &    $17.1 \pm 0.7$ & &  $31.9\pm 0.5$ \\ 
		0.75  & &   $17.4 \pm 0.6$  & &  $33.6 \pm 0.5$ \\
		1.0   & &  $18.1 \pm 0.7$ & & $35.6 \pm   0.7$ \\
		1.25  & & $19.6 \pm 0.7$ & & $36.5\pm 0.7$  \\
		1.5   & & $20.4 \pm 0.7$ & & $38.1 \pm   0.7$ \\
		1.75  & & $20.5 \pm 0.7$ & & $38.7 \pm  0.7$ \\
		2.0   & & $20.5 \pm 0.7$ & & $40.5 \pm 0.7 $ \\
		\hline
		\hline
	\end{tabular}
	\caption{Vacancy formation energy for a single vacancy $E_v$, and for two vacancies $E_{2v}$ in a fcc $^4$He crystal with $N = 108$ sites, at density $\rho=0.0294$ \AA$^{-3}$.}
	\label{table1}
\end{table}

Finally, for the specific case of the fcc crystal at the lowest density, we  
calculated the energy cost of introducing two vacancies in the system, $E_{2v}$. 
To do this, we  performed PIMC simulations of $N-2$ $^4$He atoms, in a box 
commensurate with a fcc lattice of $N$ sites. The activation energy of the 
bi-vacancy can be computed with the formula $E_{2v} = E  ( N-2 , (N-2/N)  V ) - 
(N-2) \, E(N,V)/N$ (i.e., the same as for $E_v$ in Equation (\ref{enervacan}) but changing 
$N-1$ to $N-2$). In Table \ref{table1}, we report the results obtained as a 
function of the temperature, for both $E_v$ and $E_{2v}$. The comparison between 
these two quantities can give some insight into the discussion about the stability 
of vacancies inside the bulk crystal and its possible aggregation and 
posterior \textit{evaporation}. Indeed, the condition $E_{2v} < 2 E_v$ should 
indicate the tendency for the two vacancies to approach each other, in order to 
reduce their energy cost. Although a precise estimation of the difference 
$\Delta = E_{2v} - 2 E_v$ is not feasible due to the large error bars of our 
data, we see that the activation energy of the bi-vacancy is systematically 
lower that twice the energy of a single vacancy: in particular, this effect is clearer at low temperature, where the vacancies delocalize.
\vspace{-9pt}

\section{Conclusions}
\label{sec:conclusions}

We performed  a set of PIMC simulations to investigate the behavior of point defects (vacancies) in solid samples of $^4$He at low temperatures. Our results confirm the old conjecture firstly proposed by Andreev and Lifshitz~\cite{andreev}, according to which vacancies in quantum crystals, below~a certain temperature, are delocalized over several lattice sites and can turn the system into a supersolid phase. Indeed, we notice that the behavior of the one-body density matrix of a defected crystal at high temperature is comparable with that of a perfect one. On the contrary,  at low temperature, $\rho_1(r)$ at large distances presents a plateau, which signifies a finite condensation fraction in the many-body system ($n_0 \sim 10^{-3}$).

Moreover, a clear signal of the delocalization of the vacancies can be found in 
their energetic properties. The behavior of the activation energy $E_v$ of a 
single vacancy as a function of the temperature shows a characteristic two-step 
structure indicating that, whenever the vacancies are delocalized, their~energy 
cost is reduced by an amount of few Kelvin. This behavior is independent of the 
density and the lattice geometry of the simulated crystal. Both the emergence 
of ODLRO and the two-step behavior of $E_v$ points to the existence of a 
second-order phase transition between the normal and superfluid crystals. 
However, we cannot determine the critical temperature with precision since that 
would require a finite-size scaling study, which would be extremely demanding 
in computer time, and~that it is out of the scope of the present work.

We have also calculated the energetic properties of the solid sample when two vacancies are present. The comparison between the activation energy of the bi-vacancy $E_{2v}$ with that of a single vacancy $E_v$ seems to indicate an effective attraction between the defects, for the whole range of temperatures studied here. However, it is difficult to get a deeper insight into the question about the spatial clusterization of vacancies, due to the large error bars of our estimations that hinders any firm~statement.

\vspace{6pt} 

\acknowledgments{This work was supported in part by MINECO (Spain) under Grants No. FIS2014-56257-C2-1-P and No. FIS2017-84114-C2-1-P. The authors acknowledge discussions on the vacancy issues with Y. Lutsyshyn.}


\begin{thebibliography}{999}
\bibitem{balibar_rev} Balibar, S.; Caupin, F. Supersolidity and disorder. \emph{J. Phys. Condens. Matter} \textbf{2008}, \emph{20}, 173201.

\bibitem{claudi_rmp}
Cazorla, C.; Boronat, J. Simulation and understanding of atomic and molecular quantum crystals. \emph{Rev.~Mod.~Phys.} \textbf{2017}, \emph{89}, 035003.

\bibitem{Chan1}
Kim, E.; Chan, M. Probable observation of a supersolid helium phase. \emph{Nature} \textbf{2004}, \emph{427}, 225--227.

\bibitem{Chan2}
Kim, E.; Chan, M.H. Observation of superflow in solid helium. \emph{Science} \textbf{2004}, \emph{305}, 1941--1944.

\bibitem{beamish}
Day, J.; Beamish, J. Low-temperature shear modulus changes in solid $^{4}$He and connection to supersolidity. \emph{Nature} \textbf{2007}, \emph{450}, 853--856.

\bibitem{west}
West, J.T.; Syshchenko, O.; Beamish, J.; Chan, M.H. Role of shear modulus and statistics in the supersolidity of helium. \emph{Nat. Phys.} \textbf{2009}, \emph{5}, 598--601.

\bibitem{chan_no}
Kim, D.Y.; Chan, M.H. Absence of supersolidity in solid helium in porous vycor glass. \emph{Phys. Rev. Lett.}  \textbf{2012}, \emph{109}, 155301.

\bibitem{Pederiva97}
Pederiva, F.; Chester, G.; Fantoni, S.; Reatto, L. Variational study of vacancies in solid $^{4}$He with shadow wave functions. \emph{Phys. Rev. B} \textbf{1997}, \emph{56}, 5909.


\bibitem{Chaudhuri99}
Chaudhuri, B.; Pederiva, F.; Chester, G. Monte Carlo study of vacancies in the bcc and hcp phases of $^{4}$He \emph{Phys. Rev. B} \textbf{1999}, \emph{60}, 3271--3278.

\bibitem{Galli01}
Galli, D.E.; Reatto, L. Vacancies in Solid $^{4}$He and Bose Einstein Condensation. \emph{J. Low Temp. Phys.} \textbf{2001}, \emph{124}, 197--207.

\bibitem{Galli03}
Galli, D.E.; Reatto, L. Recent progress in simulation of the ground state of many Boson systems. \emph{Mol. Phys.} \textbf{2003}, \emph{101}, 1697--1703.

\bibitem{Galli06}
Galli, D.E.; Reatto, L. Bose-Einstein Condensation of Incommensurate Solid $^{4}$He. \emph{Phys. Rev. Lett.} \textbf{2006}, \emph{96}, 165301.

\bibitem{Boninsegni06}
Boninsegni, M.; Kuklov, A.; Pollet, L.; Prokof’ev, N.; Svistunov, B.; Troyer, M. Fate of Vacancy-Induced Supersolidity in $^{4}$He. \emph{Phys. Rev. Lett.} \textbf{2006}, \emph{97}, 080401.

\bibitem{Clark08}
Clark, B.K.; Ceperley, D.M. Path integral calculations of vacancies in solid Helium. \emph{Comput. Phys. Commun.} \textbf{2008}, \emph{179}, 82--88.

\bibitem{Cazorla09}
Cazorla, C.; Astrakharchik, G.; Casulleras, J.; Boronat, J. Bose-Einstein quantum statistics and the ground state of solid $^{4}$He. \emph{New. J. Phys.} \textbf{2009}, \emph{11}, 013047.

\bibitem{Yaros10}
Lutsyshyn, Y.; Cazorla, C.; Astrakharchik, G.; Boronat, J. Properties of vacancy formation in hcp $^{4}$He crystals at zero temperature and fixed pressure. \emph{Phys. Rev. B} \textbf{2010}, \emph{82}, 180506.

\bibitem{Rota_prl} 
Rota, R.; Boronat, J. Onset Temperature of Bose-Einstein Condensation in Incommensurate Solid $^{4}$He. \emph{Phys.~Rev. Lett. }\textbf{2012}, \emph{108}, 045308.

\bibitem{calvo}
Boninsegni, M.; Kuklov, A.; Pollet, L.; Prokof’ev, N.; Svistunov, B.; Troyer, M. Luttinger liquid in the core of a screw dislocation in helium-4. \emph{Phys. Rev. Lett.} \textbf{2007}, \emph{99}, 035301. 


\bibitem{koning}
Borda, E.J.L.; Cai, W.; de Koning, M. Dislocation Structure and Mobility in hcp $^{4}$He. \emph{Phys. Rev. Lett.} \textbf{2016}, \emph{117}, 045301.


\bibitem{andreev}
Andreev, A.; Lifshitz, I. Quantum theory of crystal defects. \emph{Sov. Phys. JETP} \textbf{1969}, \emph{29}, 1107.

\bibitem{toennies}
Benedek, G.; Kalinin, A.; Nieto, P.; Toennies, J.P. Vacancy-induced flow of solid helium. \emph{Phys. Rev. B} \textbf{2016}, \emph{93},~104505.

\bibitem{cinti14}
Cinti, F.; Macrì, T.; Lechner, W.; Pupillo, G.; Pohl, T. Defect-induced supersolidity with soft-core bosons. \emph{Nat.~Commun.} \textbf{2014}, \emph{5}, 3235.


\bibitem{Rossi08}
Rossi, M.; Vitali, E.; Galli, D.; Reatto, L. Zero-point vacancies in quantum solids. \emph{J. Low Temp. Phys.} \textbf{2008}, \emph{153}, 250--265.

\bibitem{Pessoa09}
Pessoa, R.; De Koning, M.; Vitiello, S. Zero-point divacancy concentration in the shadow wave function model for solid $^{4}$He. \emph{Phys. Rev. B} \textbf{2009}, \emph{80}, 172302.

\bibitem{Lutsyshyn10}
Lutsyshyn, Y.; Cazorla, C.; Boronat, J. Instability of Vacancy Clusters in Solid $^{4}$He.  \emph{J. Low Temp. Phys.} \textbf{2010}, \emph{158}, 608--614.

\bibitem{CeperleyRev}
Ceperley, D.M. Path integrals in the theory of condensed helium. \emph{Rev. Mod. Phys.} \textbf{1995}, \emph{67}, 279.

\bibitem{Sakkos09}
Sakkos, K.; Casulleras, J.; Boronat, J. High order Chin actions in path integral Monte Carlo. \emph{J. Chem. Phys.} \textbf{2009}, \emph{130}, 204109.

\bibitem{BoninsegniWorm}
Boninsegni, M.; Prokof’ev, N.; Svistunov, B. Worm algorithm and diagrammatic Monte Carlo: A new approach to continuous-space path integral Monte Carlo simulations. \emph{Phys. Rev. E} \textbf{2006}, \emph{74}, 036701.

\bibitem{sarsa00}
Sarsa, A.; Schmidt, K.; Magro, W. A path integral ground state method. \emph{J. Chem. Phys.} \textbf{2000}, \emph{113}, 1366--1371.

\bibitem{rossi09}
Rossi, M.; Nava, M.; Reatto, L.; Galli, D. Exact ground state Monte Carlo method for Bosons without importance sampling. \emph{J. Chem. Phys.} \textbf{2009}, \emph{131}, 154108.

\bibitem{rota_pre}
Rota, R.; Casulleras, J.; Mazzanti, F.; Boronat, J. High-order time expansion path integral ground state. \emph{Phys.~Rev.~E} \textbf{2010}, \emph{81}, 016707.

\bibitem{aziz}
Aziz, R.A.; McCourt, F.R.; Wong, C.C. A new determination of the ground state interatomic potential for He$_{2}$. \emph{Mol. Phys.} \textbf{1987}, \emph{61}, 1487--1511.


\bibitem{rota_jltp}
Rota, R.; Boronat, J. Condensate Fraction in Liquid $^{4}$He at Zero Temperature. \emph{J. Low. Temp. Phys.} \textbf{2012}, \emph{166},~21--32.

\bibitem{Diallo12}
Diallo, S.; Azuah, R.T.; Abernathy, D.L.; Rota, R.; Boronat, J.; Glyde, H.R. Bose-Einstein condensation in liquid $^{4}$He near the liquid-solid transition line. \emph{Phys. Rev. B} \textbf{2012}, \emph{85}, 140505.

\bibitem{gillan}
Gillan, M. Calculation of the vacancy formation energy in aluminium. \emph{J. Phys. Condens. Matter} \textbf{1989}, \emph{1}, 689--711.




\end{thebibliography}
\end{document}